\documentclass[sn-mathphys,Numbered]{sn-jnl}% Math and Physical Sciences Reference Style

\usepackage{graphicx}%
\usepackage{multirow}%
\usepackage{amsmath,amssymb,amsfonts}%
\usepackage{amsthm}%
\usepackage{mathrsfs}%
\usepackage[title]{appendix}%
\usepackage{xcolor}%
\usepackage{textcomp}%
\usepackage{manyfoot}%
\usepackage{booktabs}%
\usepackage{algorithm}%
\usepackage{algorithmicx}%
\usepackage{algpseudocode}%
\usepackage{listings}%
\usepackage{epigraph}
\newcommand{\beq}{\begin{equation}}
\newcommand{\eeq}{\end{equation}}
\newcommand{\bea}{\begin{eqnarray}}
\newcommand{\eea}{\end{eqnarray}}

\usepackage{ulem}
\providecolor{added}{rgb}{0.0,0.5,0.0}
\providecolor{deleted}{rgb}{1,0,0}
\newcommand{\added}[1]{{\color{added}{}#1}}

\begin{document}

\title[Rising Tides] {Rising Tides: Analytic Modeling of Tidal Effects in Binary Neutron Star Mergers
%\epigraph{The rising tide lifts all the boats.}{\textit{John F. Kennedy}}
}
\author[1]{\fnm{Alexander} \sur{O'Dell}}
\equalcont{Undergraduate student} %{These authors contributed equally to this work.}
\author*[1]{\fnm{Maria C} \sur{Babiuc Hamilton}}\email{babiuc@marshall.edu}

\affil[1]{\orgdiv{Department of Physics}, \orgname{Marshall University}, \orgaddress{\street{1 John Marshall Drive}, \city{Huntington}, \postcode{25755}, \state{WV}, \country{USA}}}
%\affil[]{something}

%%==================================%%

\abstract{
%Purpose:
%\deleted{Methods:}
%Results:
%Conclusion
%  
% \epigraph{Unforgettable words}{Peter W.}
%
The gravitational waves produced by binary neutron star mergers offer a unique window into matter behavior under extreme conditions.
In this context, we model analytically the effect of matter on the gravitational waves from binary neutron star mergers.  
We start with a binary black hole system, leveraging the post-Newtonian formalism for the inspiral and the Backwards-one-Body model for the merger.
We combine the two methods to generate a baseline waveform and we validate our results against numerical relativity simulations. 
Next, we integrate tidal effects in phase and amplitude to account for matter and spacetime interaction, by using the NRTidal model, and test its accuracy against numerical relativity predictions, for two equations of state, finding a mismatch around the merger. 
Subsequently, we lift the restriction on the coefficients to be independent of the tidal deformability, and recalibrate them using the numerical relativity predictions.  
We obtain better fits for phase and amplitude around the merger, and are able to extend the phase modeling beyond the merger.
We implement our method in a new open-source Python code, steered by a Jupyter Notebook.
Our research offers new perspectives on analytically modeling the effect of tides on the gravitational waves from binary neutron star mergers.
}
%%================================%%
% As a guide the abstract should not exceed 200 words. 




\maketitle

%%================================%%
%%===FIRST REVIEW===%
%\added{}
%\deleted{}

\section{Introduction}\label{intro}
Neutron stars are extremely dense remnants of massive stars, at the brink of collapsing into a black hole, with masses comparable to that of the sun, contained in only twenty kilometers diameter. 
They are captivating celestial objects, under the action of powerful gravitational and magnetic fields, which cannot be replicated on Earth.
These exceptional properties present them as excellent astrophysical laboratories for investigating the behavior of matter in extreme conditions of density, pressure and temperature.
When two neutron stars collide and merge, they emit gravitational waves, accompanied by electromagnetic radiation, matter, and neutrinos; carrying valuable information about their masses, sizes and interior structure. 

The first direct detection of a gravitational wave (GW) signal from a binary neutron star (BNS) collision, named GW170817 \cite{arXiv:1710.05832}, was accompanied by a gamma-ray burst \cite{arXiv:1710.05834} and ignited a new explosion, called a kilonova, powered by the radioactive decay of the heavy elements synthesized in the merger \cite{arXiv:1710.05843}. 
This twin detection marked the onset of the golden era in neutron star research, alluding to the precious metals such as gold synthesized and expelled during the collision, and has inspired intense theoretical investigations.
These studies provided us with a wealth of new insights into the yet unknown internal structure of the neutron stars \cite{arXiv:1805.11581, arXiv:1805.11963, arXiv:1807.06437}, the characteristics of their magnetic fields \cite{arXiv:2004.10105}, and the outflow of heavy matter during collision \cite{arXiv:1809.06843}. 

%Motivation
Unfortunately, simultaneous detections of gravitational waves and light are few and far between. 
Out of the almost 100 such events reported to date \cite{arXiv:2105.09151}, the overwhelming majority came from binary black hole (BBH) collisions and only one other, named GW190425, was produced by an unusually heavy BNS collision \cite{arXiv:2001.01761, arXiv:2001.04502}. 
The two BNS mergers detected so far raised many questions, proving again that nature is more complicated than our models \cite{arXiv:2005.02964, arXiv:2107.08053}. 
It is thus imperative to revise our current understanding of the physics involved in those models, in order to gain insight on BNS mergers.
This is a timely topic, because in the next decade hundreds of such GW events might be detected \cite{arXiv:1902.09485}, with the advent of the new generation of gravitational wave observatories, including both ground-based detectors such as Cosmic Explorer \cite{arXiv:1907.04833} and the Einstein Telescope \cite{arXiv:1912.02622}, and space-borne instruments such as LISA cite{arXiv:2001.09793}, DECIGO \cite{arXiv:2006.13545} and TianQin \cite{arXiv:1512.02076}.

Efforts in the analytical modeling of the tidal interactions during BNS mergers have focused on developing post-Newtonian (pN) approximations for the late-inspiral phase \cite{arXiv:1009.4919, arXiv:1205.3403, arXiv:1412.4553,  arXiv:1812.02744}, relying on the effective-one-body (EOB) approach \cite{arXiv:gr-qc/9811091, arXiv:0911.5041, arXiv:1602.00599, arXiv:1806.01772}. 
As those approximations become less accurate near merger \cite{arXiv:1310.8288, arXiv:1702.02053, arXiv:2009.08467}, alternative methods have emerged, notably the closed-form tidal approximants \cite{arXiv:1706.02969, arXiv:1802.06518, arXiv:1804.02235, arXiv:1905.06011, arXiv:1910.08971, arXiv:2205.06023}, which combines pN, tidal EOB, and numerical relativity data.

In this study, we employ the NRTidal approximant \cite{arXiv:1706.02969, arXiv:1804.02235, arXiv:1905.06011}, an elegant model that adjusts the analytically calculated binary black hole (BBH) waveforms by adding a closed-form expression to account for the tidal influences in the phase and amplitude of the GW.  
This model was used for estimating source properties \cite{arXiv:2108.01045} and constraining the equation of state for ultra-dense matter in the first two BNS detections \cite{arXiv:2210.09259}, and it is implemented in the LIGO Scientific and Virgo Collaborations Algorithm Library Suite (LALSuite) \cite{LALSuite}.

Our goal is to analytically model the effect of tides on the GW signal during the inspiral and to accurately extend it through the merger of the BNS systems considered. 
We aim to create an open-source, easily replicable code that generates comprehensive analytical templates for gravitational radiation during BNS collisions. By doing so, we facilitate independent consistency checks and assist in defining the domains of validity for the approximation models employed. Ultimately, our work contributes to the development of a universal analytical model encapsulating the BNS dynamics and the effect of tides on GW emission.

This paper unfolds as follows: first, we present the point-particle system that characterizes BBH collisions, and calculate the GW by combining the post-Newtonian formalism for inspiral evolution with the Backwards-one-Body model for the merger.
As baseline GW we use the fully analytical model for BBH collision we developed in \cite{arXiv:1810.06160} and expanded in \cite{arXiv:2203.08998}, based on \cite{arXiv:1609.05933} for the inspiral and \cite{arXiv:1810.00040} for the merger. 
We prove our model's validity and efficiency through a comparison with numerical relativity (NR), utilizing the \texttt{SXS:BBH:0180} template from the Simulating eXtreme Spacetimes (SXS) catalogue \cite{arXiv:1904.04831}. 
We then incorporate the tidal deformability into the point-particle waveform, both as polynomial corrections and as rational functions approximations to the phase and amplitude, by using the coefficients proposed in \cite{arXiv:1905.06011, arXiv:1804.02235}.
Next, we test our implementation against \texttt{SXS:NSNS:0001/0002}  for two equations of state \cite{arXiv:1812.06988}.

Finally, we push this model past the merger, by performing a new fit to the numerical BNS data for the tidal phase and amplitude, from which we derive updated values for the polynomial coefficients, expanding thus its applicability.
Upon determining the new coefficients, we reconstruct the tidal correction in phase and amplitude beyond the merger and analyze the tidal influence on the early stages of the collision, revealing the effect of the matter interaction on the system's orbits. 

In this work we consider two neutron stars with total mass $M = M_A + M_B$, $M_A \le M_B$ and mass ratio $q = M_B/M_A \ge 1$. 
We express time, space and energy in geometric units (with $G=c=1$), in terms of the binary mass $M$, written as a multiple of the sun mass $M_{\odot}$.
 
%%================================%%
\section{The Baseline Model}\label{BBHmodel}
We start with assembling the baseline analytical model for the calculation of the GWs from a BBH collision, by following the common procedure as we detailed in \cite{arXiv:1810.06160, arXiv:2203.08998}. 
First, we split the binary motion in two regions: the weak field, during the inspiral, and the strong field, during the merger, and apply different mathematical formalisms to obtain the waveform for each region. 
We then generate the complete GW template for the whole binary evolution by matching those two regions in frequency around the last stable orbit, and building the hybrid waveform. % close to the merger, with a step function. 
Lastly, we compare our model against a numerically generated equal mass BBH collision, to uphold its validity.

%%================================%%
\subsection{The Baseline Inspiral Model}\label{BBHinspiral}
Let us consider a tight binary system of separation $r$, in quasi-circular orbit. 
By defining the reduced mass $\mu = (M_A M_B)/M$ and the symmetric mass ratio $\eta= \mu/M$, we reduce it further to a single particle of mass $\mu$ and position $r$, orbiting around the mass $M$ located at the center of mass.
We require the system to dissipate GWs and are led to the balance equation:
\beq
F(t) = -\frac{dE(t)}{dt}.
\label{eq:dE}
\eeq
This formula states that the GW flux $F(t)$ is emitted at the expense of the orbital energy $E(t)$, causing the orbit to shrink.
Eq.(\ref{eq:dE}) can be rewritten in terms of a small factor $x_{pN}=(v/c)^2$ called post-Newtonian parameter, where $v$ is the orbital velocity of the binary and $c$ is the speed of light (see \cite{arXiv:0907.3596, arXiv:1102.5192, arXiv:1310.1528, arXiv:1902.09801}).
\beq
\frac{dx_{pN}(t)}{dt} = -\frac{F(t)}{dE(t)/dx_{pN}(t)}.
\label{eq:dxpNE}
\eeq
Using the post-Newtonian (pN) approximation, we expand the deviation from Newtonian gravity as a perturbation in power series of $x_{pN}$.
From the many different methods of solving eq.(\ref{eq:dxpNE}), in this work we chose the TaylorT4 approximant, shown to agree best with numerical simulations \cite{arXiv:1705.07089}. 
Within this method, eq.(\ref{eq:dxpNE}) becomes \cite{arXiv:1609.05933,arXiv:1810.06160}: 
\beq
\frac{dx_{pN}(t)}{dt}\bigg |^{\rm \tfrac{N}{2}} 
= \frac{x^{5}_{pN}(t)}{M} \sum_{j=0}^{N} \xi_{j} x^{j/2}_{pN}(t).
\label{eq:dxpN}
\eeq
where $N/2$ denotes the pN expansion order. 
In this work we go up to $3.5$ in the leading pN order, adding self-force and hereditary correction terms up to $6$pN order, to increase the accuracy in modeling the region near the merger \cite{arXiv:1609.05933}.
By integrating eq.(\ref{eq:dxpN}) with the coefficients $\xi_{j}$ we obtain the evolution of the pN parameter $x_{pN}$.  %given in Appendix \ref{appendixA}, 
Next, we use Kepler's third law $v^2 =  (M \Omega (t))^{2/3}$ where $\Omega$ is the angular orbital velocity, to obtain the equation for the orbital phase: 
\beq
\frac{d \Phi_{pN}(t)}{dt} = \Omega(t) =\frac{ x_{pN}(t)^{3/2}}{M}.
\label{eq:dphipN}
\eeq
We readily integrate eq.(\ref{eq:dphipN}) to find the time evolution of the orbital inspiral phase.

We reuse Kepler's third law written as: $v = \sqrt{M/r}$ to extract the orbital separation as $r = M/x_{pN}$, which we then expand in powers series of $x_{pN}$ for increased accuracy:
\beq
r_{pN}(t) = \frac{M}{x_{pN}(t)} \sum_{j=0}^{3} \rho_j x_{pN}(t)^{j} .
\label{eq:rpN}
\eeq
%and we give the coefficients $ \rho_{j}$ in Appendix \ref{appendixB}. 

Lastly, we calculate the GW amplitude for optimal orientation of the source, 
\bea
\label{eq:Ains}
& A_{\cal R}(t) =  - 2 \frac{\mu }{R} \left [\frac{M}{r_{pN}(t)} + r_{pN}(t)^2 \left(\frac{d \Phi_{pN}(t)}{dt}\right)^2 - \left(\frac{d r_{pN}(t)}{dt}\right)^2 \right ], 
\nonumber \\
& A_{\cal I}(t) = - 4 \frac{\mu}{R} r_{pN}(t) \frac{d \Phi_{pN}(t)}{dt} \frac{d r_{pN}(t)}{dt} .
\eea 
Now we have all the pieces necessary to construct the dimensionless strain, defined as:
\beq
\label{eq:hGW}
h_{pN}(t) = A_{pN}(t) e^{-2\mathrm{i} \phi_{pN}(t)} 
%= A_{pN}(t) \sin{(2\phi_{pN})} -\mathrm{i}  A_{pN}(t) \cos{(2\phi_{pN})}
\eeq
Mathematically, the strain is decomposed into two transverse \emph{quadrupolar} polarization modes, with $h_+$ representing the real part of eq.(\ref{eq:hGW}) and $h_\times$ the imaginary part,
\bea
\label{eq:hpcGW}
&h_{+,pN}(t) = 
A_{\cal R}(t) \cos{(2\phi_{pN})} + \mathrm{i}  A_{\cal I}(t) \sin{(2\phi_{pN})},
\nonumber \\
&h_{\times,pN}(t) = 
A_{\cal R}(t) \sin{(2\phi_{pN})} - \mathrm{i}  A_{\cal I}(t) \cos{(2\phi_{pN})}.
\eea
This is because the GWs compress the space in one direction while simultaneously stretching it in the orthogonal direction, such that the signal goes twice through maxima and minima during one orbital cycle, making the frequency of the GW twice the orbital frequency. 
This technique, albeit powerful, is valid only if the gravitational field is sufficiently weak and the orbital velocity is smaller than the speed of light. 

%%================================%%
\subsection{The Baseline Merger Model}\label{BBHmerger}
Going beyond the pN approximation brings us up against the strong gravitational field around the merger.
The transition between the weak and strong field is marked by the innermost stable circular orbit (ISCO), defined as the last stable orbit a
particle would have when orbiting around a black hole. 
When the masses of the celestial objects are comparable, this location is not well defined \cite{arXiv:gr-qc/0703053}, but can be approximated with the last stable photon orbit, called the light-ring (LR), which is close to the peak of the curvature potential \cite{arXiv:1609.00083}.
From there on, through the merger and ringdown, we use the {\em Backwards-One-Body} formalism \cite{arXiv:1810.00040}.
This technique starts from the perturbed final black hole resulting after collision and builds the GW signal back in time to the end of the inspiral, assuming we know the mass, spin and ringdown frequency of the remnant.
In this model, the strain of the GW signal can be modeled analytically as exponentially decaying sinusoids that break free from the LR on null geodesics \cite{arXiv:0905.2975}:
\beq
h_{\mathit{BoB}}(t) = \sum_{lmn} A_{\mathit{lmn}}e^{ \mathrm{i} \omega_{\mathit{lmn}}t} e^{-t/\tau_{\mathit{lmn}}},
\eeq
where $l$ is the principal, $|m| \le l $ the azimuthal and $n$ is the overtone index of a mode. 
This kind of perturbation is called \emph{quasinormal} (QNM) ringing with frequency $\omega_{lmn}$ and damping time $\tau_{lmn}$.
The frequency of the lower mode is ($l=2, m=1, n=0$) coincide with the orbital frequency and we will denote it as $\Omega_{QNM}$. 
The ($l=2, m=2, n=0$) mode can be derived from this mode with the simple relation $\omega_{22} =  2 \Omega_{QNM}$.
This mode carries away most of the GW's energy ($\approx 95\%$) \cite{arXiv:gr-qc/0512160}, while the higher harmonics, being much quieter, can be usually neglected.
We will drop the $(l,m,n)$ indices and consider that the strain is well described by the dominant mode.

The BoB model requires the QNM frequency and damping time, which are determined by the spin and mass of the final black hole.
We will estimate the final spin of the resulting black hole with a polynomial of coefficients $s_{ij}$, as given in \cite{arXiv:0710.3345, arXiv:0904.2577}:
\beq
\chi_{f} =\sum^3_{i,j=0} s_{ij} \eta^i\chi^j_{\mathit{eff}}.
\label{eq:chifin}
\eeq
We define an effective spin:
\beq
\chi_{\mathit{eff}}= \frac{M_A^2 \chi_A+M_B^2 \chi_B}{M_A^2+M_B^2},
\eeq
with $\chi_{\mathit{A,B}}= S_{\mathit{A,B}}/M_{\mathit{A,B}}^2$ the dimensionless individual spins and $S_{A,B}$ the spin angular momentum of each black hole entering the merger. 
In our calculations we pick $\chi_{A,B} =0$, as in the NR simulations we use for comparison, although the mean spin for low-spinning, astronomical neutron stars is $\chi_{A,B} \approx 2 \times 10^{-3}$ \cite{arXiv:1102.1500}.
%We give the coefficients $s_{ij}$ in Appendix \ref{appendixC}. 

For the final mass we use the fit to NR given in \cite{arXiv:1312.5775}  for comparable mass binaries, 
\beq
M_{f}= M (1  - \tilde E_{\mathit{GW}}) - M_{\mathit{disk}}
\label{eq:Mf}
\eeq
where $\tilde E_{\mathit{GW}} = E_0 +  E_2 \chi_f^2 + E_4 \chi_f^4$ is the dimensionless energy released in GWs. 
For the disk mass we take as upper limit of $M_{\mathit{disk}}\approx 10^{-2}M$ \cite{arXiv:1908.02350}.
The coefficients for $E_i$ are given in Table III of \cite{arXiv:1312.5775}.

Next, we calculate the dominant resonant frequency with a polynomial fit:
\beq
M_f \Omega_{\mathit{QNM}} = f_1 + f_2(1 - \chi_{f})^{f_3},
\label{eq:Om}
\eeq
and use a similar formula for the quality factor:
\beq
Q = q_1 + q_2(1 - \chi_{f})^{q_3}. 
\label{eq:Q}
\eeq
The pairs $(f_{i}, q_{i})$ are taken from Table VIII of \cite{arXiv:gr-qc/0512160}.
Lastly, the damping time is:
\beq
\tau_{\mathit{QNM}} = \frac{2 Q}{\omega_{\mathit{QNM}}}.
\label{eq:tau}
\eeq
%For convenience, we include the coefficients $E_i$, $(f_{i}$ and $q_{i})$ in Appendix \ref{appendixC}.

The orbital angular frequency $\Omega_{\mathit{BoB}}(t)$ is given in this model by:
\beq
\label{eq:omgBoB}
\Omega_{\mathit{BoB}}(t) = \left( \Omega_i^4+\kappa \left[ \tanh \left(\frac{t-t_0}{\tau}\right) - \tanh \left(\frac{t_i-t_0}{\tau}\right) \right]\right)^{1/4}.
\eeq
Here, $\Omega_i$ is the initial frequency, $t_0$ is the time at which the strain of the GW reaches its peak amplitude, and $t_i$ the initial time , marking the transition between the weak and strong regime. 
The parameter $\kappa$ in eq.(\ref{eq:omgBoB}) assures the continuity between the end of the inspiral and the beginning of the merger, and is given by:
\beq
\label{eq:constk}
\kappa = \left[ \frac{\Omega_{\mathit{QNM}}^4-\Omega_i^4}{1 -\tanh\left(\frac{t_i-t_0}{\tau} \right)}\right].
\eeq

The essential variable in the BoB model is the initial time $t_i$, that locks in the frequency $\Omega_i$ at the beginning of the merger to the frequency at end of the inspiral:
\beq
t_i = t_0 - \frac{\tau}{2}\ln \left( \frac{ \Omega_{\mathit{QNM}}^4 - \Omega_i^4}{2\tau\Omega_i^3\dot{\Omega}_i} - 1 \right).
%\label{eq:tisimp}
\eeq

We obtain the phase $\Phi_{\mathit{BoB}}(t)$ by integrating eq.(\ref{eq:omgBoB}) between $t_i$ and a final time $t_f$:
\beq
\Phi_{\mathit{BoB}}(t) = \int_{t_i}^{t_f} \Omega_{\mathit{BoB}}(t) dt.
%\label{eq:phiBoB} 
\eeq

The amplitude of the GW signal is modeled with the simple function:
\beq
A_{\mathit{BoB}}(t) = A_0 \mbox{sech} \left(\frac{t-t_0}{\tau}\right),
\label{eq:ampBoB}
\eeq
where $A_{0}$ is a scaling factor. 
This amplitude is  taken to correspond to the Weyl scalar $|\Psi_4(t)|$, which is related to the strain by the formula
 \beq
\label{eq:Psi4}
\Psi_4(t) = \frac{\partial^2 h_{\mathit{GW}}(t)}{\partial^2 t}.
\eeq
The model rests upon the assumption that the amplitude changes much slower compared to the phase during the merger, and uses the simple expression for the strain:
\beq
h_{\mathit{BoB}}(t) = -\frac{A_{\mathit{BoB}}(t)}{\omega_{\mathit{BoB}}(t)^2}e^{-2\mathrm{i}\Phi_{\mathit{BoB}}(t)},
\label{eq:strainBoB}
\eeq
where $\omega_{\mathit{BoB}}(t) = 2\Omega_{\mathit{BoB}}(t)$ is the frequency of the dominant mode. 
Please note that in this model the influence of the proportionality coefficients arising from the double integration of eq.(\ref{eq:Psi4}) over time is overlooked. 
In reality, both the magnitude and the peak location of the amplitude are affected by this integration. 
As a result, we expect this model to yield a slight time deviation in predicting the merger position.
We assessed the strengths and weaknesses of the BoB model in a previous work \cite{arXiv:2205.14742}.

%%================================%%
\subsection{The Complete Baseline Model}\label{BBHhybrid}
Let us now assemble the two forms of the GW strain by gluing  the waveform around the LR, at the transition between the weak and strong field, estimated by \cite{arXiv:gr-qc/0001013}:
\beq
r_{\mathit{LR}} = 2 M \left [1 + \cos{\left(\frac{2}{3}\cos^{-1}(-\chi_{f})\right)}\right].
\label{eq:rLR}
\eeq
The frequency and phase matching is robust against slight variations around this location, while the continuity of the amplitude is best ensured by choosing $r_{LR}\approx4 M$, representing the location of the retrograde photon orbiting a maximally rotating BH.

We stitch the amplitudes of the two models together by first normalizing the BoB amplitude with its peak value and then rescaling the pN amplitude with the normalized BoB amplitude at $t_i$, by using the formula:
\beq
\bar A_{\mathit{pN}}= A_{\mathit{pN}} \frac{\bar A_{\mathit{BoB}}(t_i)}{A_{\mathit{pN}}(t_{t})},
\eeq
where the quantities with an overhead bar represent their rescaled values.
We build the complete hybrid model by gluing together the waveforms of the two domains at $t_i$ with the following simple piecewise (step) function:
\beq
\sigma(t)= \left\{
  \begin{array}{lr} 
      0 & t < t_i \\
      1 & t \ge t_i 
      \end{array}
\right.
\eeq
For example, the hybrid orbital frequency of the GW has the form:
\beq
\Omega_{\mathit{hyb}}(t)=(1 - \sigma(t)) \Omega_{pN}(t - t_t) + \sigma(t)\Omega_{\mathit{BoB}}(t-t_i),
\eeq
We form the hybrid amplitude, phase and strain of the GW with the same technique. 

Before moving forward, we must test the accuracy of our implementation with a numerical waveform. 
We chose \texttt{SXS:BBH:0180}, corresponding to an equal-mass, non-spinning BBH configuration of normalized mass and initial separation of $18 M$ in geometric units \cite{arXiv:1904.04831}. %r_\mathit{NR, 0} = 
We use the same configuration, but start our analytic model at a separation of $15 M$, which for a regular equal-mass BNS with a total mass of $2.8M$ translates to a  distance of abut $45 \mathrm{km}$ between the centers of mass. %a bit earlier, at $r_\mathit{hyb, 0} = 19M$.
We start the BoB model with the initial orbital frequency ($\Omega_i$, $\dot{\Omega}_i$) of the pN model at the end of the inspiral marked by the LR, and build the hybrid baseline model. 
Next, we interpolate NR data onto a time array of constant time steps, as used by the analytical model, and align the strain at the beginning. 
This alignment starts at the initial frequency of the analytical model and extends over a duration equal to the initial complete period of the strain. %\cite{arXiv:1904.04831}.
%$M\omega_0 = 0.0245746$ 

\begin{figure}[ht]
\centering
\includegraphics[width=1.0\textwidth]{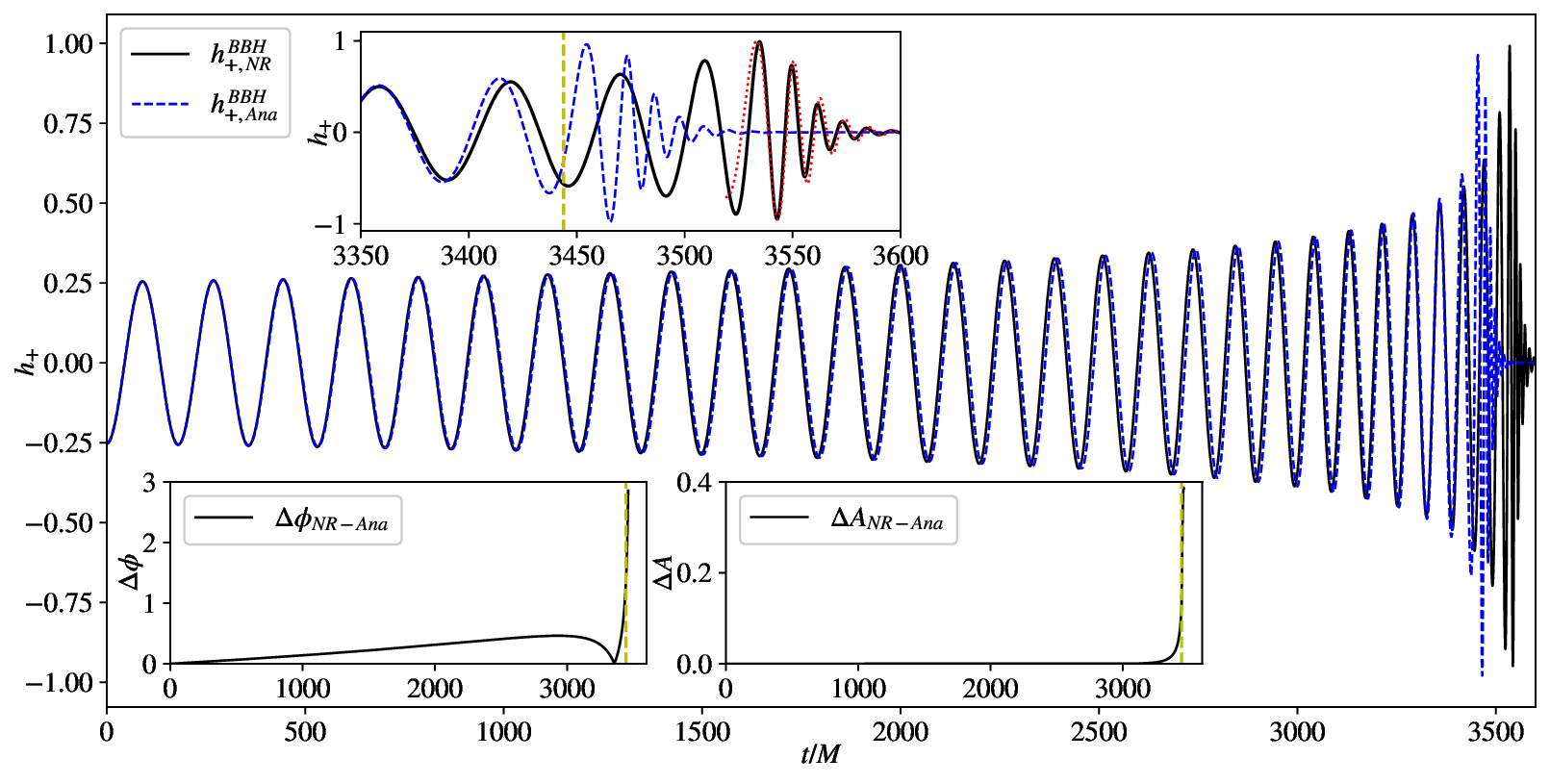}
\caption{Comparison between the numerical and the analytic plus component of the strain, matched over a period around the initial frequency. The upper inset shows that the strain peaks are slightly displaced, but the merger is well modeled if we overlap them (dotted red curve). The dashed vertical yellow line marks the LR.
The two lower insets contain the difference in phase (lower left) and amplitude (lower right) between our model and \texttt{SXS:BBH:0180}. Only expected small differences before the merger are visible.}
\label{Figure1}
\end{figure}

We plot in Fig.~\ref{Figure1} a comparison between the evolution of the $h_+$ strain component obtained with our hybrid model, and the $+$ polarization of the GW strain from the NR simulation, as well as the differences in phase and amplitude (lower insets). 
We observe a significant overlap between the strain from the analytical model and the NR simulation throughout the entire inspiral, extending up to the LR. The faster merger occurring in the analytical model is primarily attributed to the Post-Newtonian (PN) approximation's limited reach into the strong field domain.
In subsequent studies we aim to refine this model by including higher-order post-Newtonian (pN) terms as they become available.
The shift in the position of the merger is accentuated by the BoB model because, as we explained earlier, it assumes a simplified expression for the amplitude. We analyzed this deviation from the NR predicted waveform at the merger in \cite{arXiv:2205.14742}, and in future work we intend to correct the merger amplitude by incorporating the neglected integration coefficients.
The phase concordance between the analytical model and the NR data is consistent throughout the entire waveform during the inspiral phase, remaining within the sub-radian accuracy of the BBH numerical waveforms \cite{arXiv:1812.06988}.
We see an increase in the phase difference above $1$ \texttt{rad} only at the LR and reaching $2.8$ \texttt{rad} by the time of merger, which is within the numerical BNS dephasing due to tidal disruptions \cite{arXiv:1812.06988}. 
Similarly, the amplitude correlates well up to the LR, with the discrepancy becoming more pronounced as the merger approaches.
The addition of tidal effects to this analytic model will overwrite the errors around the merger, because a BNS system reaches merger earlier than its corresponding BBH.
%Keep in mind that a BNS system reaches merger earlier than its corresponding BBH, due to  
%the tidal effects that influence both the phase and amplitude of the GW near the merger. 
%Therefore, 

%%================================%%
\section{The BNS Model}\label{BNSmodel}
Thus far, we have not taken into consideration the internal structure of the neutron stars, which is determined by the equation of state
and accounts for the effect of the tidal interactions.
It is expected for all neutron stars in the universe to be described by a unique equation of state, but its expression is not currently known, only estimated by theory \cite{arXiv:1904.04233} and experiments \cite{arXiv:1904.08907} in nuclear physics.  
The tidal effects become important in the late stage of the inspiral, inducing deformations in the stars and driving them to merge faster, thus affecting the emitted GW signal both in phase and in amplitude. 
It is this effect that will allow us to determine the equation of state corresponding to the dense matter inside neutron stars from the direct observations of the amplitude and phase of the gravitational waves emitted during BNS collisions.

We can encode the tidal effects into the tidal correction strain:
\beq
h_{T}(t) = A_{T}(t) e^{-2\mathrm{i} \phi_{T}(t)} 
\label{eq:htide}
\eeq
where $A_{T}$ and $\phi_{T}$ are the analytic tidal corrections to the GW amplitude and phase. 
We find the analytical GW strain of the BNS model by multiplying eq.(\ref{eq:htide}) to the strain of the baseline model \added{\cite{arXiv:1804.02235}}: 
\beq
h_{\mathit{BNS}}(t) = A_{\mathit{BNS}}(t) e^{-2\mathrm{i} \phi_{\mathit{BNS}}(t)} = h_{\mathit{BBH}}(t) h_{T}(t) = A_{\mathit{BBH}} A_{T}(t) e^{-2\mathrm{i} (\phi_{BBH}(t)+ \phi_{T}(t))}.
\label{eq:hBNS}
\eeq
This equation tells us that we must add the tidal phase correction to the baseline, 
\beq
\phi_{\mathit{BNS}}(t) = \phi_{\mathit{BBH}}(t) + \phi_{T}(t),
\label{eq:phiBNS}
\eeq
while for the amplitude, we should multiply the tidal correction to the baseline:
\beq
%\log_{10} A_{\mathit{BNS}}(t) = \log_{10}A_{\mathit{BBH}}(t) + \log_{10}A_{T}(t).
A_{\mathit{BNS}}(t)=A_{\mathit{BBH}}(t)  A_{T}(t).
\label{eq:AmpBNS}
\eeq
 We implement the tidal effects as polynomial functions of the variable $x = (M\Omega_{hyb})^{2/3}$ and build the analytical expressions for the BNS phase and amplitude provided by the pN and NRTidal models, then we validate our implementation by comparing it two open-source BNS simulations from the SXS catalogue \cite{arXiv:1812.06988} .

%%================================%%
\subsection{The BNS Tidal Phase}\label{BNSphase}
As the stars get closer together, their tidal deformability increases, affecting the binding orbital energy. 
Looking back to eq.(\ref{eq:dE}), we should not be surprised to see that this will change the flux of gravitational waves emitted as the orbit shrinks. 
%Again, we express the contribution of the tides to the orbital energy as expansions in term of the $x_{\mathit{pN}}$ factor. 
The tidal correction becomes significant only as the stars approach the merger, not entering in the pN decomposition until the $5^{th}$ pN correction term \cite{arXiv:0709.1915}. 
The correction to the GW phase induced by the tidal deformation of one star due to the gravitational pull produced by its companion is known up to $7.5$-pN \cite{arXiv:1203.4352}, and can be written as:
\beq
\phi_{T} = \frac{13  x^{2.5}}{8 \eta} \kappa_{ A, \mathit{eff}} \left( 1 + c_{ A,1} x + c_{ A,1.5} x^{1.5} + c_{ A ,2} x^2 + c_{ A, 2.5} x^{2.5} \right ) + \left [ A \leftrightarrow B \right ]
\label{eq:phiT}
\eeq
with $x$ from the baseline BBH model and the pN tidal phase coefficients $c_{\left [ A , B \right ],i}$ depending on the mass ratio, as given in \cite{arXiv:1905.06011}. 
The tidal effects are encoded in $\kappa_{\left [ A , B \right ],\mathit{eff}}$, named the effective tidal coupling constant, introduced in \cite{arXiv:1706.02969}:
\beq
\kappa_{\mathit{eff}} = \kappa_{A, \mathit{eff}} + \kappa_{B, \mathit{eff}}
= \frac{3}{32} \left (\tilde \Lambda_A + \tilde \Lambda_B \right )
=  \frac{3}{16} \tilde \Lambda,
\eeq
where:
\beq
\tilde \Lambda_A= \frac{16}{13}\frac{(M_A + 12 M_B) M_A^4 \Lambda_A}{M^5} ;~~
\tilde \Lambda_B= \frac{16}{13}\frac{(M_B + 12 M_A) M_B^4 \Lambda_B}{M^5} 
\eeq
are the symmetric mass-weighted tidal deformabilities \cite{arXiv:1310.8288}, defined in terms of $\Lambda_{\left [ A, B \right ]}$, the dimensionless tidal deformability of each star in the binary \cite{arXiv:0711.2420}. 
Let us take a closer look at this quantity, because it is its dependence on the equation of state and mass that can reveal the internal structure of the neutron stars in the binary by measuring the phase of their GW emission and fitting it with analytical or numerical templates. 
In general, the nuclear models give numerical values for the dimensionless tidal deformability between $10^2$ and $10^3$, varying inversely proportional with the mass and the compactness of the star \cite{arXiv:1808.02858}. 
So called hard nuclear equations of state models describe less compact stars, predicting higher values for the tidal deformability, while nuclear models with soft equations of state favor more compact stars and tidal deformabilities towards lower values. 
The dimensionless tidal deformability calculated from the GW170817 event gave, for a neutron star of mass $1.4 M_{\odot}$, a value somewhere in the middle range, namely $\tilde \Lambda \approx 400$ \cite{arXiv:1805.11579}.
 
We now return to the tidal correction to the GW phase, and, as mentioned before, choose the NRTidal model for supplying the expression of the tidal phase evolution. 
This analytical approximation, first introduced in \cite{arXiv:1706.02969} and subsequently improved in \cite{arXiv:1804.02235, arXiv:1905.06011}, is calibrated to NR, and because of the scarce availability of numerical simulations of BNS systems with unequal mass ratio, includes only non-spinning, equal-mass configurations.
In this case, eq.(\ref{eq:phiT}) simplifies to a polynomial of constant coefficients: 
\beq
\phi_{T,\mathit{eq}} = \frac{13  x^{2.5}}{8 \eta} \kappa_{\mathit{eff}} \left ( 1 + \sum_{i=2}^{N}c_{i/2} x^{i/2}  \right )
= \frac{13  x^{2.5}}{8 \eta} \kappa_{ \mathit{eff}} P(x).
\label{eq:phiTeq}
\eeq
In order to fit eq.(\ref{eq:phiTeq}) to NR data, NRTidal replaces the polynomial $P(x)$ with a rational function $R(x)$ given by a Pad\'e approximant of constant coefficients: 
\beq
R(x) = \frac{1 + n_1 x + n_{1.5}x^{1.5} + n_2 x^2 + n_{2.5} x^{2.5} + n_3 x^3}
	                  {1 + d_1 x + d_{1.5}x^{1.5} + d_2 x^2},
\label{eq:Pade}
\eeq
determined from the NR, and restricted to enforce consistency with the analytical coefficients entering the polynomial $P(x)$.

Using one set of coefficients for $c_{i}$ and two sets of coefficients for $(n_i, d_i)$ from \cite{arXiv:1706.02969, arXiv:1804.02235, arXiv:1905.06011}, we implement the tidal correction to our hybrid baseline model as: %listed in Appendix \ref{appendixD} for convenience, 
\beq
\phi_{T,\mathit{pN}} =  \frac{13  x^{2.5}}{8 \eta} \kappa_{\mathit{eff}} P(x); ~~
\phi_{T,\mathit{F1(F2)}} =  \frac{13  x^{2.5}}{8 \eta} \kappa_{\mathit{eff}} R(x).
\label{eq:phiTeqS}
\eeq
Note that, in contrast to the corresponding equations presented in \cite{arXiv:1706.02969, arXiv:1804.02235, arXiv:1905.06011}, we found necessary to use a positive sign in eq.(\ref{eq:phiTeqS}), in order to obtain the expected behavior for the BNS phase evolution when we use eq.(\ref{eq:phiBNS}) to obtain the analytic BNS phase. %e correction to the baseline BBH phase. 
%of a BNS system 
%\beq
%\phi_{\mathit{BNS}} = \phi_{\mathit{BBH}} + \phi_{T}.
%\eeq

It is time to check our implementation and test how well does the NRTidal model predict the tidal phase against numerical simulations.
We choose two equal-mass, non-spinning BNS systems with two different equations of state \cite{arXiv:1812.06988}, publicly available in the SXS GW catalogue \cite{arXiv:1904.04831}.
The first numerical GW used is \texttt{SXS:NSNS:0001}, with mass $M_1 = 2.8$ and an ideal gas equation of state of polytropic index $\Gamma = 2$ corresponding to a dimensionless tidal deformability $\tilde \Lambda_{\Gamma2} = 791$. 
The second is \texttt{SXS:NSNS:0002}, with mass $M_2 = 2.7$ and a piecewise polytropic equation of state for cold dense matter called MS1b \cite{arXiv:0812.2163}, with dimensionless tidal deformability $\tilde \Lambda_{\mathit{MS1b}} = 1540$.
While these tidal deformabilities surpass the one estimated from the GW170817 observation, they are not yet ruled out and are useful as an upper bound. 
We also note that in the pN and NRTidal models the tidal correction to the phase depends only linearly on the dimensionless tidal deformability, and the coefficients should remain independent of it. 

We start with the assumption that at a considerable distance from the merger, the tidal effects are negligible and the BNS signal is virtually indistinguishable from that of the BBH. 
This allows us to rescale the phase and shift the time of the numerical BNS simulations to start from zero at the reference frequency of each simulation, marking when the initial burst of spurious transient radiation has dissipated in the NR simulation \cite{arXiv:1904.04831}.
Then, we find the time and phase corresponding to the same frequency in the numeric BBH model and align the numerical BNS waveforms with the corresponding values of the numeric BBH model.
We take the difference in phase between the numeric BNS and BBH data, up to the merger frequency, taken where the numeric BNS amplitude reaches its peak.

\begin{figure}[ht]
\centering
\includegraphics[width=1.0\textwidth]{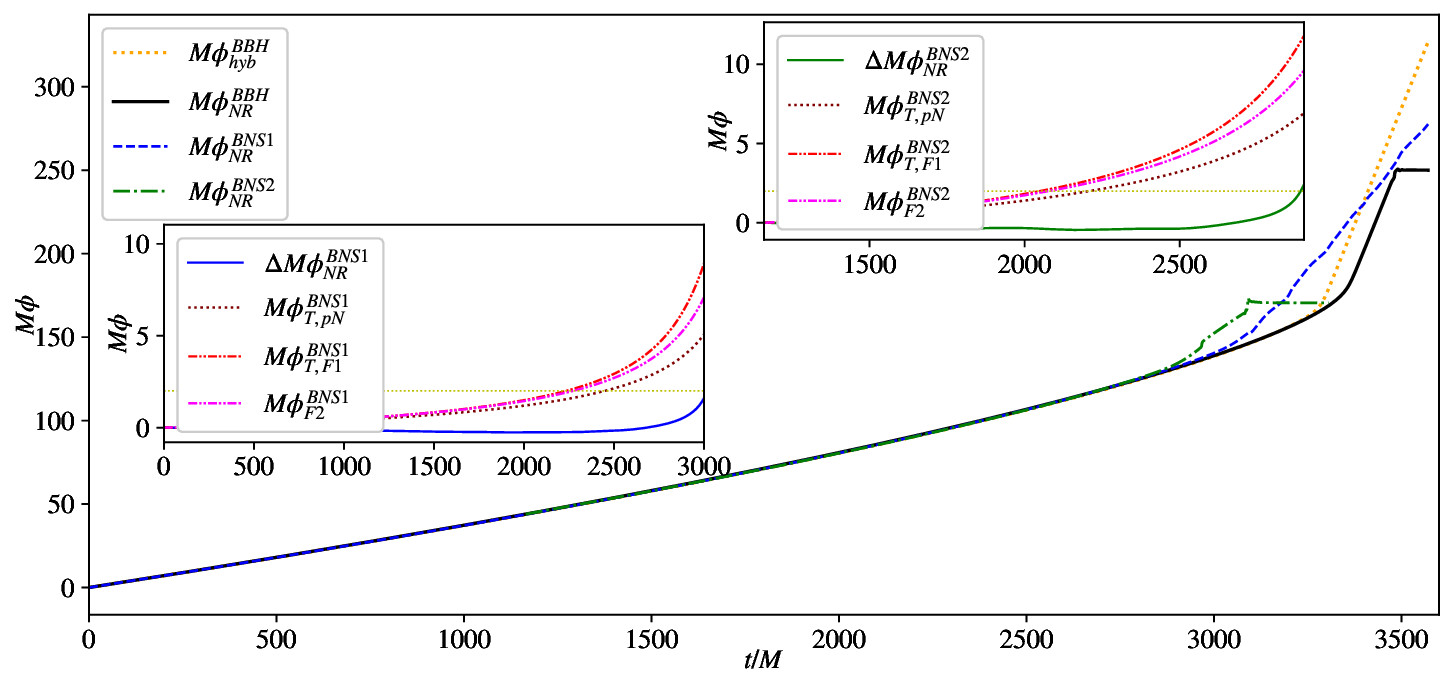}
\caption{Phase comparison between the baseline model (solid orange), \texttt{SXS:BBH:0180} (solid black), \texttt{SXS:NSNS:0001} (dashed blue) and \texttt{SXS:NSNS:0002} (dash-dot green) overlapped at the reference frequencies of the NR simulations. 
The insets contain our analytical BNS phase calculated with the pN (dotted maroon) and NRTidal (dash-dot-dot red and magenta) corrections,  plotted up to the merger against the numerical phase difference for the two cases. The horizontal dotted yellow curve marks the maximum uncertainty in phase of $2$ \texttt{rad}.}\label{Figure2}
\end{figure}

We plot in Fig.~\ref{Figure2} the phase comparison between our analytic baseline hybrid model, \texttt{SXS:BBH:0180}, \texttt{SXS:NSNS:0001} (BNS1) and \texttt{SXS:NSNS:0002} (BNS2). 
%We confirm that the steep increase in the BNS phase marking the merger happens faster, thus the small mismatch in merger time we uncovered in our baseline model does not affect the analytical model for the BNS phase. 
The insets in Fig.~\ref{Figure2} contain the analytical models for the BNS tidal phase $\phi_{T,\mathit{pN}}$ and $\phi_{T,\mathit{NRfit}}$ from eq.(\ref{eq:phiTeqS}), compared to the numerical BNS tidal phase.
We mark with a dashed yellow curve the phase of $2$ \texttt{rad}, representing the upper limit of uncertainty in phase due to numeric errors \cite{arXiv:1812.06988}. The maximum modeling error of our baseline BBH phase in the domain considered is below $0.5$ \texttt{rad}.
We observe that the NRTidal phase (dash-dot-dot red and magenta) slightly overestimates both the pN (dotted maroon) and the NR phase, and is more accurate for the BNS1 case, corresponding to the smaller tidal deformabiliy. 
This suggests that the assumption of the coefficients being independent of the tidal deformability may be too restrictive.
In Section \ref{results} we will delve into the behavior of the phase around the merger and propose new coefficients for the tidal correction to push the model past the merger.  

%%================================%%
\subsection{The BNS Tidal Amplitude}\label{BNSamp}
\begin{figure}[ht]
\centering
\includegraphics[width=1.0\textwidth]{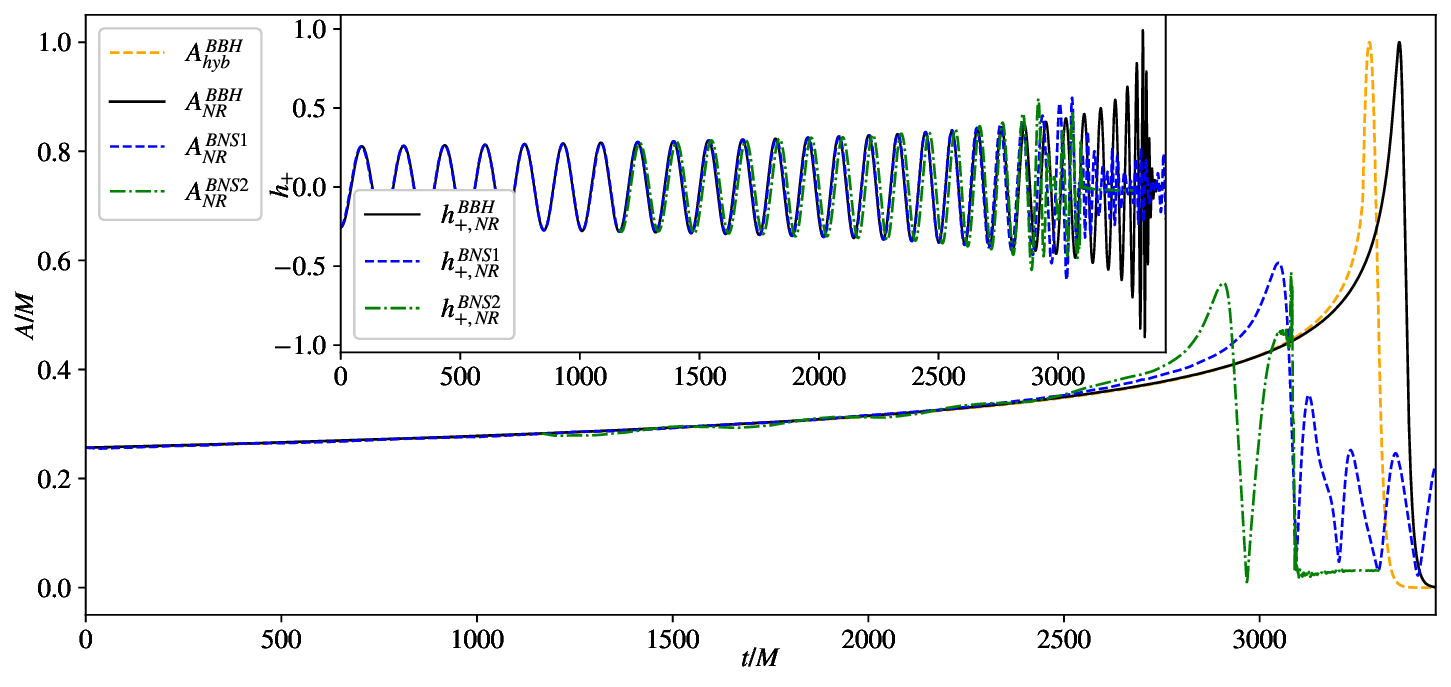}
\caption{Amplitude comparison between \texttt{SXS:NSNS:0001} (dashed blue) and \texttt{SXS:NSNS:0002} (dash-dot green) rescaled at the reference frequencies of the NR simulations. 
The inset contain a comparison of the numerical $h_{+}$ between the BNS and BBH data.}
\label{Figure3}
\end{figure}

The amplitude of the GW is extremely small, and its increase is less dramatic than the phase accumulation observed up to the merger, defined as the point of maximum amplitude \cite{arXiv:2006.03168}. 
This makes the task of accurately modeling analytic tidal contributions to the amplitude more difficult compared to the phase.
Taking advantage of the relatively small change in the GW amplitude, we can effectively use a low-order polynomial within the context of the pN approximation, of the form \cite{arXiv:1203.4352, arXiv:1905.06011}:
\beq
	A_{T}(x) = \frac{8 M \eta}{D_L}\sqrt{\frac{\pi}{5}}x^6 
						\kappa_{\mathit{eff},A}(\hat c_{A, 0}+ \hat c_{A, 1} x)
						 + \left [ A \leftrightarrow B \right ],
\label{eq:AmpTpN}	
\eeq
where $D_L$ is the distance from the detector to the BNS system and $\hat c_{\left [A ,B \right ],i}$ are the pN tidal amplitude coefficients, given in \cite{arXiv:1203.4352, arXiv:1905.06011}.
Let's assume again a similar-mass binary, where only the symmetric mass ratio $\eta$ retains the information on the mass ratio. 
Now, the pN tidal correction in amplitude eq.(\ref{eq:AmpTpN}) simplifies to:
\beq
	A_{T,\mathit{pN}}(x) = A_{T,\mathit{eq}}(x)  = \frac{M \eta }{21 D_L}\sqrt{\frac{\pi}{5}}x^6 \kappa_{\mathit{eff}} (672 -11 x).
\label{eq:AmpTpNeq}
\eeq
As expected, this approximation becomes less reliable as the binary approaches the merger.
The NRTidal model corrects it by adding a dependence of $x$:
\beq
	A_{T,d}(x) = \frac{A_{T,\mathit{pN}}(x)}{1+d x},~
\label{eq:AmpTd}
\eeq
where the parameter $d$ is fixed by the identity \cite{arXiv:1905.06011}:
\beq
	d = \frac{1}{x} \left( \frac{A_{T, pN}(x)}{A_{\mathit{mrg}}}-1\right)\bigg|_{x=x_{\mathit{mrg}}}.
\label{eq:dAmpNR}
\eeq
We compute the merger amplitude $A_{\mathit{mrg}}$ using an analytically predicted quasi-universal relation
valid at the moment of merger \cite{arXiv:1402.6244},
that provides the peak amplitude only as function of the tidal coupling constant $\kappa_{\mathit{eff}}$ \cite{arXiv:1905.06011}:
\beq
	A_{\mathit{mrg}} = \frac{M \eta}{D_L}\frac{1.6498 (1+2.5603 \times 10^{-2}\kappa_{\mathit{eff}} - 1.024 \times 10^{-5}\kappa_{\mathit{eff}}^2)}{1 + 4.7278 \times 10^{-2}\kappa_{\mathit{eff}}}.
\eeq
As a cautionary remark, eq.(\ref{eq:dAmpNR}) requires that we provide an analytical expression for $x_{\mathit{mrg}}$ as well, without relying on numerical data. 
A good approximation is to assume that the stars merge when they come in contact, and $x_{\mathit{mrg}}$ is given by:
\beq
x_c = \frac{M}{R_A + R_B},
\label{eq:xc}
\eeq
where $R_{A,B}$ are the radii of the stars, for which we use the value of $11.5 \texttt{km}$ as specified in \cite{arXiv:1812.06988}.
Please note that this simplified expression does not account for the deformations of the stars.
To address this, we use an alternative route, and calculate $x_{\mathit{mrg}} = (M\Omega_{\mathit{mrg}})^{2/3}$ from the analytic expression for the merger frequency given in \cite{arXiv:1810.03936}:
\beq
M\Omega_{\mathit{mrg}} = \sqrt{q} \frac{ 0.178 (1+3.354 \times 10^{-2}\kappa_{\mathit{eff}} + 4.315 \times 10^{-5}\kappa_{\mathit{eff}}^2)}{1 + 7.542 \times 10^{-2}\kappa_{\mathit{eff}} + 2.236 \times 10^{-4}\kappa_{\mathit{eff}}^2)}.
\label{eq:MOmrg}
\eeq
We chose to use the result calculated with eq.(\ref{eq:MOmrg}) in our implementation.
Working within the same assumption that far from the merger the BNS and BBH waveforms have indistinguishable amplitudes, we rescale the numeric BNS amplitude to coincide with the numeric BBH amplitude for a binary with the same total mass at the reference frequency.
We align the BNS and BBH amplitudes and plot them in Fig.~\ref{Figure3}. 
This is not fulfilled exactly, as we see in the inset of Fig.~\ref{Figure3}, where we plot a comparison of their $h_{+}$ strain because at the BNS reference frequency, tidal effects are already present in the NR waveforms.

Let us now add the tidal amplitude correction to our baseline model, using eq.(\ref{eq:AmpBNS}).
We assume that the amplitudes are indistinguishable at the reference frequency of the numerical BNS data, although the waveform is already in the regime where tidal effects are non-negligible. Nonetheless, we consider this error in amplitude as not essential for our analysis and consider that $A_{T, \mathit{ref}} \approx  1$.
%This equation can also be written as: $A_{\mathit{BNS}}(t) = A_{\mathit{BBH}}(t) A_{T}(t)$, 
We capture this behavior for the evolution of our analytic BNS amplitude, by changing $A_{T}$ to:
%by pulling it to $0$ at the reference frequency and adding 1 such that:
\beq
	A_{T}(x) \rightarrow 1 + A_{T}(x).
\label{eq:AmpTcorr}	
\eeq

\begin{figure}[ht]
\centering
\includegraphics[width=1.0\textwidth]{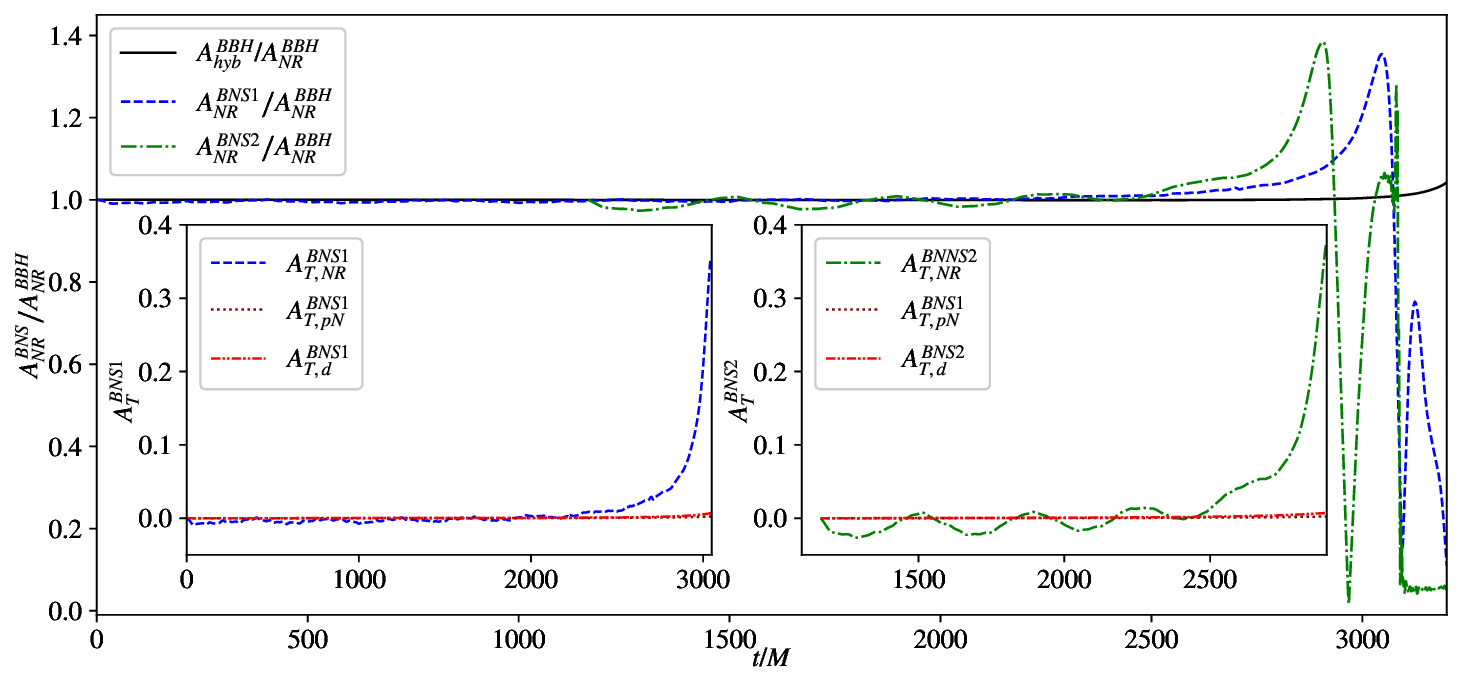}
\caption{Amplitude comparison between the baseline model (solid black), \texttt{SXS:NSNS:0001} (dashed blue) and \texttt{SXS:NSNS:0002} (dash-dot green) overlapped at the reference frequencies of the NR simulations and divided by the amplitude for SXS:BBH:0180. 
The insets contain the amplitudes for the pN (dotted maroon) and NRTidal (dash-dot-dot red) approximations,  plotted up to the merger against the numerical BNS amplitude for the two cases.}\label{Figure4}
\end{figure}

We plot in Fig.~\ref{Figure4} the ratio of the amplitudes for \texttt{SXS:NSNS:0001} (BNS1) and \texttt{SXS:NSNS:0002} (BNS2) to the numeric BBH amplitude compared with the ratio of our hybrid analytic baseline amplitude o the numeric BBH amplitude. 
The insets in Fig.~\ref{Figure4} contain the analytical approximations for the BNS tidal amplitude, as in eq.(\ref{eq:AmpTpNeq}) and eq.(\ref{eq:AmpTd}), up to the corresponding BNS merger frequency.
We observe that both analytical approximations underestimate the steep increase in amplitude around the merger. 
We will return to this comparison in Section \ref{results}, when we will make a new fit for the analytical amplitude to the numerical data and taper the amplitude past the merger using a Hanning window. 

%%================================%%
\section{Modeling the BNS Merger}\label{results}
 Up to this point we have limited ourselves with modeling the tidal effects up to the merger, and confirmed that both the pN and NRTidal approximations hold reasonably well in comparison with the two numerical relativity simulations considered, although they were not included in the calibrations \cite{arXiv:1706.02969, arXiv:1804.02235, arXiv:1905.06011}.
Let us now take a closer look at the behavior of the analytical models for the tidal phase and amplitude as we approach the merger. 
Indeed, we shall see a noticeable mismatch with the numerical simulations, and will attempt to improve the model by performing our own fit only up to the merger.
We will see that because of the smooth evolution of the phase and amplitude even after the moment of the merger, we can use the new  coefficients obtained from our fit to push the model past the location where the stars touch.
We succeed in extending the model for the phase beyond the merger, and devise a method to determine how far we can reach, where we end it with a taper and continue with the baseline model.
We carry the amplitude up to the merger and terminate it with a Hann taper, to ensure a smooth and continuous transition into the post merger. 

%%================================%%
\subsection{New Fit for the Tidal Phase} 
\begin{figure}[ht]
\centering
\includegraphics[width=1.0\textwidth]{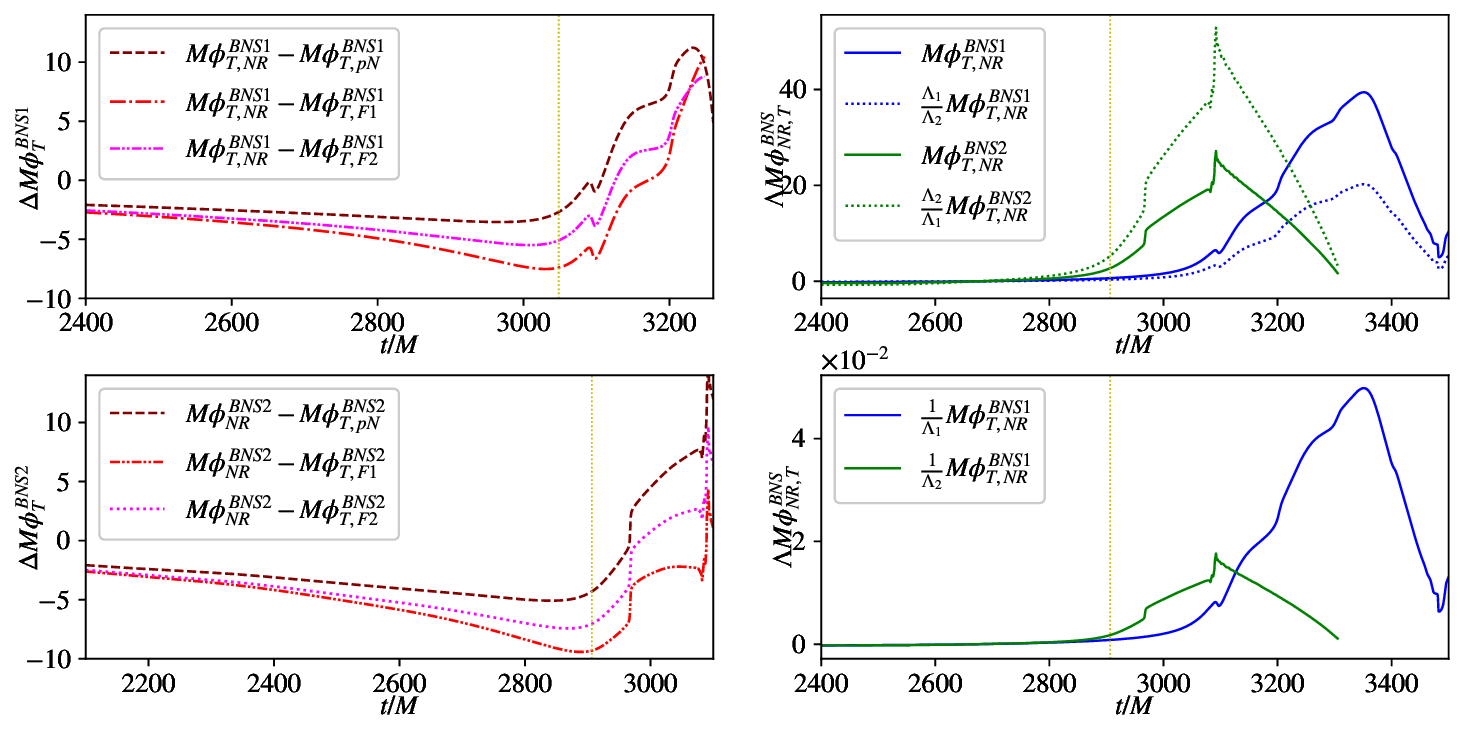}
\caption{Left plots: Difference between the numerical tidal phase correction and the analytic pN and the three NRTidal models (pN, $F1$ and $F2$) for the tidal phase for the two BNS systems, close to the merger, and extended beyond it.  Right plots: true and rescaled tidal phase for the two BNS systems, overlapped at the merger time.
The vertical yellow dotted line marks the merger.}
\label{Figure5}
\end{figure}

Let us proceed by first taking the difference between the numerical BNS and BBH phase for the two system considered.
This is the {\em true} tidal correction up to $2$\texttt{rad.} numerical error \cite{arXiv:1812.06988}) that we subsequently use to compare with the analytical models for the tidal phase. 
We plot in the left side of Fig. \ref{Figure5} the difference between the numerical tidal phase and the three analytical approximations for $\phi_{T, pN}$, $\phi_{T, F1}$ and $\phi_{T, F2}$ with the coefficients from \cite{arXiv:1706.02969, arXiv:1804.02235, arXiv:1905.06011}.
We start at a time about 1000 M before the merger, where the differences between the analytic and numerical phase become noticeable.
We then extend the model past the merger, until the approximation exhibits a sharp increase and breaks down. 
We observe an increased difference between the analytical and numerical tidal phase near the merger, above the uncertainty of $2$\texttt{rad.}
%\added{After the merger} \deleted{T}\added{t}
In the right side of Fig. \ref{Figure5} we plot the true tidal phase (upper plot), as well as the scaled tidal deformability (lower plot).
The true tidal phase scales with the tidal deformability before the merger, as expected, but after the merger is larger for the first (BNS1) system compared to the second (BNS2), appearing to scale inversely with the tidal deformability. 
The analytical phase, which depends linearly on the tidal deformability, doesn't provide an accurate estimation after the merger. Specifically, the postmerger tidal phase is underestimated for the first system and overestimated for the second. 
%This does make sense, considering that the postmerger of the BNS system with softer equation of state 

We assume that the Pad\'e approximant from eq.(\ref{eq:Pade}) is complex enough to model the smooth increase of the tidal phase at the merger, and proceed with performing new curve fits for the analytical tidal phase to the numerical tidal phase, scaled by the tidal deformability.
We use as initial guesses the coefficients from the pN \cite{arXiv:1706.02969}, $F1$ \cite{arXiv:1804.02235} and $F2$ \cite{arXiv:1905.06011} NRTidal model and stop the fit at the merger. 
Irrespective of the initial fitting coefficients, we obtain comparable sets of new coefficients for the polynomial modeling the tidal interaction of a BNS system with a given deformability.
However, in contrast to the original fits, we do not find any longer a one-size-fits-all set of coefficients, namely our coefficients do depend on the equation of state considered. 

To select the best fit among the three variations of the analytical model, we calculate and compare the sum of squared residuals (SSR) for each numerical dataset. 
For BNS1, we select new parameters obtained by fitting to the numerical data, using the $F2$ coefficients as the initial guess.
For BNS2, we choose the parameters derived from fitting the numerical data with the $F1$ coefficients as initial guess.
Subsequently, we compare the residuals for the chosen coefficients between the two datasets and select the optimal fitting parameters for BNS1 as the new coefficients.

We give in Table \ref{tab1} the average values of the new coefficients obtained, in comparison with the $F2$ NRTidal coefficients. 
\begin{table}[h]
\caption{New merger fit for tidal phase, dependent on the tidal deformability.}\label{tab1}%
\begin{tabular}{@{}lll@{}}
\toprule
Coefficient & Original $F2$ fit \cite{arXiv:1905.06011} & New fit \\
\midrule
$n_1$      & $-15.2452$ & $-214.95$  \\
$n_{1.5}$ & $31.5423$  & $2016.23$ \\
$n_{2}$    & $-80.9260$ & $-7743.40$ \\
$n_{2.5}$ & $312.482$  & $13772.45$ \\
$n_{3}$    & $-342.155$ & $-9381.96$ \\
$d_1$      & $-20.2372$ & $-42.62$ \\
$d_{1.5}$ & $39.3962$  & $151.06$ \\
$d_{2}$    & $-5.36163$ & $-150.08$ \\
\botrule
\end{tabular}
\end{table}
%%%%
We need to alert readers that the large differences between our new fitting coefficients and the original ones are primarily due to our efforts to model the highly nonlinear tidal interactions during the merger, which is a step beyond the bounds of the analytical approximation
The assumptions made in the original model that the coefficients were independent of the equation of state, are no longer applicable.

We use the new fitting coefficients to reconstruct the analytical tidal phase, then we incorporate it into the baseline framework, thus obtaining a new analytical model for the BNS phase that extends beyond the merger.
Following this, we develop a procedure that determines the termination point of our phase, where we apply a Heaviside function, allowing only the phase of the baseline model to continue from then on.
First, we calculate the difference between our new analytical tidal phase and the true tidal phase. After that, we determine the time derivative of this difference.
Next, we identify potential cutoff points at locations where this derivative switches sign. 
Lastly, we select the final cutoff point as the last point for which the deviation from the numerical phase is less than $3\%$, to optimize the fit with the true tidal phase.
Here we terminate our analytical tidal phase with a Heaviside function and continue smoothly with the baseline phase, obtaining a complete phase representation.

\begin{figure}[ht]
\centering
\includegraphics[width=1.0\textwidth]{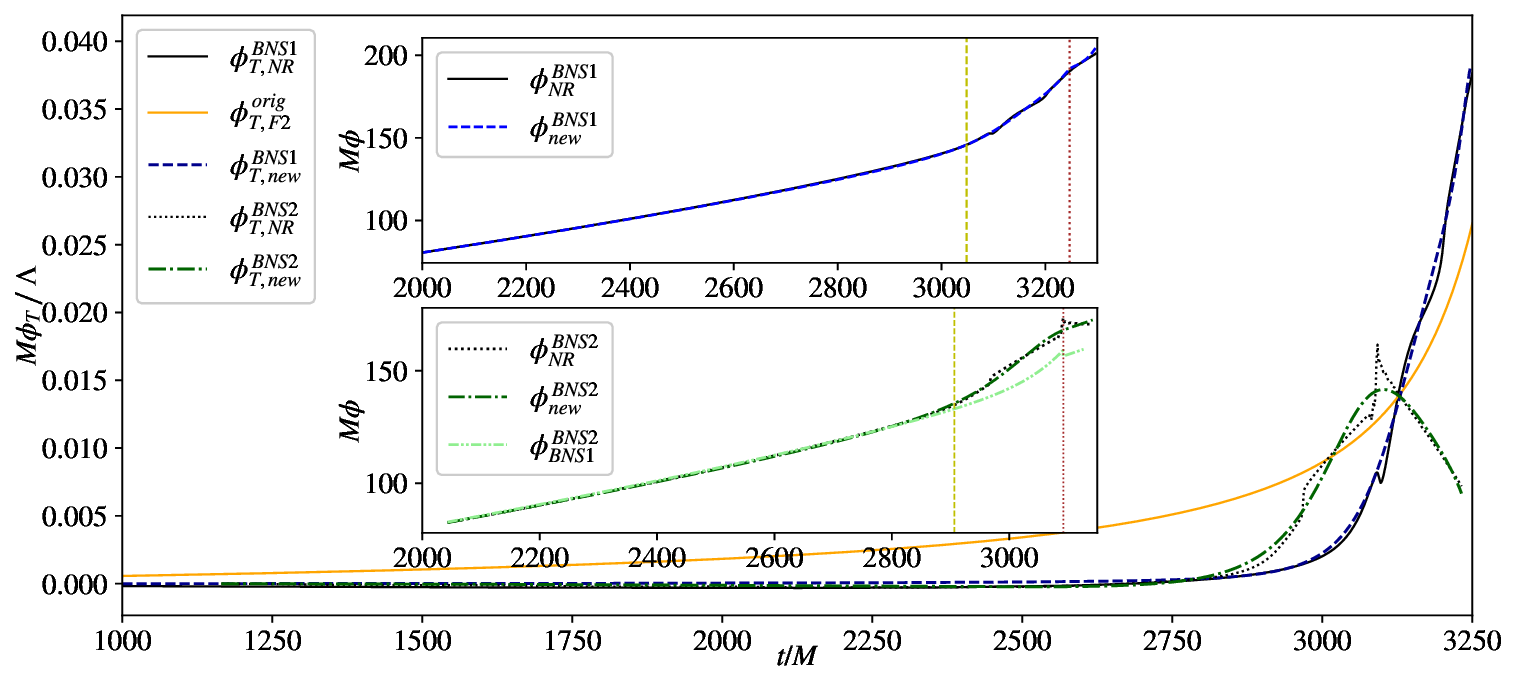}
\caption{Comparison between our analytic new fit to the tidal phase (dashed blue and dash-dot green) to the numerical tidal phase (solid and dotted black), as well as the original $F2$ fit (solid orange), for the two BNS systems, up to the merger, then continued beyond the merger.
The insets contain the reconstructed BNS phase for the two BNS systems, extended beyond the merger, as well as the new fit with the chosen coefficients to BNS2 (dash-dot-dot light green). The vertical yellow dotted line marks the merger, and the vertical brown dotted line delineates the end point where tapered.}
\label{Figure6}
\end{figure}

We plot in Fig.\ref{Figure6} our newly modeled phase. 
In the main plot we show our new analytic fit for the BNS phase tailored to the two BNS systems, against the true tidal phase.
Although we applied the curve fit only up to the merger, we see that it is able to accurately follow the numerical tidal phase beyond the merger.
We include in the two insets a comparison between our reassembled tidal phase for the two BNS systems, with the Heaviside taper applied at the cutoff point, and the numerical BNS phase. 

%the assumption of the coefficients being independent of the tidal deformability may be too restrictive.

%%================================%%
\subsection{New Fit for the Tidal Amplitude}
We use a similar procedure to model the analytical amplitude at the merger and start by calculating the numerical tidal amplitude as the ratio between the numerical BNS and BBH amplitudes for the two system considered.
This is the true tidal amplitude that we use next to compare with the analytical approximations $A_{T,\mathit{pN}}$ and $A_{T,d}$. 
We attempt to improve the amplitude modeling by applying a curve fit for eqs.(\ref{eq:AmpTpNeq}, \ref{eq:AmpTd}) to the true tidal amplitude up to the merger, using as initial guess the coefficients given in \cite{arXiv:1905.06011}. 
Again, we obtain similar sets of new coefficients regardless of the model we start with,  but depending on the value of the tidal deformability.
We apply again the same procedure of calculating and comparing the sum of squared residuals (SSR) and select the optimal fitting parameters for BNS1 amplitude as the new coefficients.

We give in Table \ref{tab2} the values of the new coefficients resulting from the new tidal amplitude fit, in comparison with the $A_{T,d}$ NRTidal coefficients.  
\begin{table}[h]
\caption{New merger fit for tidal amplitude, dependent on the tidal deformability}\label{tab2}%
\begin{tabular}{@{}lll@{}}
\toprule
Coefficient & Original fit $A_{T,d}$ \cite{arXiv:1905.06011} & New fit \\
\midrule
$d$      & $-5.96$ & $-1.3176$  \\
$p$      & $1$       & $0.1318$  \\
\botrule
\end{tabular}
\end{table}

With these coefficients, we proceed to reconstruct the new analytical amplitude. 
This time we cannot track the amplitude beyond the merger, due to its unphysical steep increase.
While the model is precise enough to capture the sharp rise in amplitude before the merger, it is too basic to accurately follow its evolution beyond the peak point. 
To complete the waveform beyond the merger, we taper the amplitude at the merger with a Hann window that terminates to zero at the phase cutoff time. 

\begin{figure}[ht]
\centering
\includegraphics[width=1.0\textwidth]{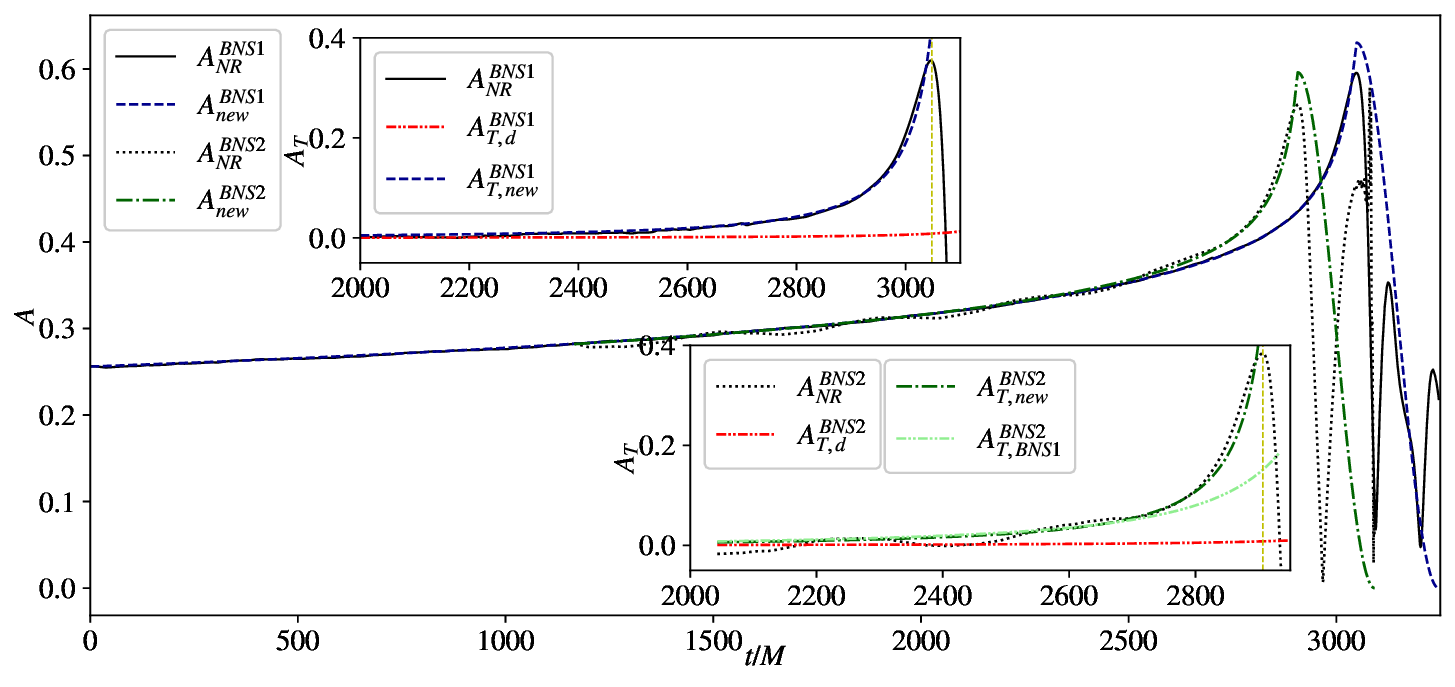}
\caption{Comparison between the two numerical BNS amplitudes (solid and dotted black) and our new fit for the amplitudes (dashed blue and dash-dot green) up to the merger, then tapered by a Hann function between merger and the cutoff time for the phase. 
The insets include our fit to the numerical tidal amplitude, compared with $A_{T,d}$. The vertical yellow dotted line marks the merger.}
\label{Figure7}
\end{figure}

We plot in Fig. \ref{Figure7} the comparison between the numerical BNS amplitude and our new amplitude with tides.
The two insets display our fit to the numerical tidal amplitude, contrasted with the pN and NRTidal tidal amplitudes for the two BNS systems.

%%================================%%
\subsection{Complete Analytic BNS Waveforms}

With these new fits obtained for the phase and amplitude, we build the complete analytical GW strain for the BNS merger using eq.(\ref{eq:hBNS}).
Although we overlooked the amplitude complexity after the merger, we modeled the phase accurately up to the cutoff time. 
This allows us to recompute the orbital frequency $\Omega_{\mathit{BNS}}$ by taking the time derivative of the phase $\phi_{\mathit{BNS}}$.
Then we recalculate $x_{\mathit{BNS}}$ and follow with the radial separation $r_{\mathit{BNS}}$, using eq.(\ref{eq:rpN}).
We also reevaluate the orbital and radial velocities and compare their evolution through the merger. 
Looking at their ratio we observe that even at the merger, where the radial velocity reaches its maximum, it represents at most $ 6\%$ of the orbital velocity, after which it falls abruptly.
In contrast, the orbital velocity keeps increasing even after the merger, reaching its peak soon thereafter, after which it starts decreasing as well.
The domain between merger and cutoff time is very short, extending up to $240 M$ for BNS1 and $180M$ for BNS2.
For the systems considered, this is between $2~\mathtt{msec}$ and $3~\mathtt{msec}$, not enough to inform us on the fate of the remnant. 
Most likely, we reach the early stages of a rapidly-rotating, tidally deformed remnant.
The GW170817 event indicates that a remnant with a total mass of $2.8 M_{\odot}$ will settle as a neutron star, the collapse to a black hole being less likely \cite{arXiv:1805.11581}.

Note that our model does not account for the mass ejected due to the tidal interactions during coalescence, or the dynamically ejected mass during the collision-induced shock of the neutron star crust when the stars come in contact. 
The matter expulsion at the contact interface is called shock ejecta, and represents a significant source of ejecta for systems with similar masses, as considered in this work. 
Additionally, our model excludes the neutrino-driven wind ejecta emitted by the HMNS remnant as it cools down.
The different types of ejecta produced during the BNS evolution form a rotating disk surrounding the remnant. 
The total mass ejected by the GW170817 event was estimated to be about $0.04 M_{\odot}$, and this is the value we consider in our work. 

\begin{figure}[ht]
\centering
\includegraphics[width=1.0\textwidth]{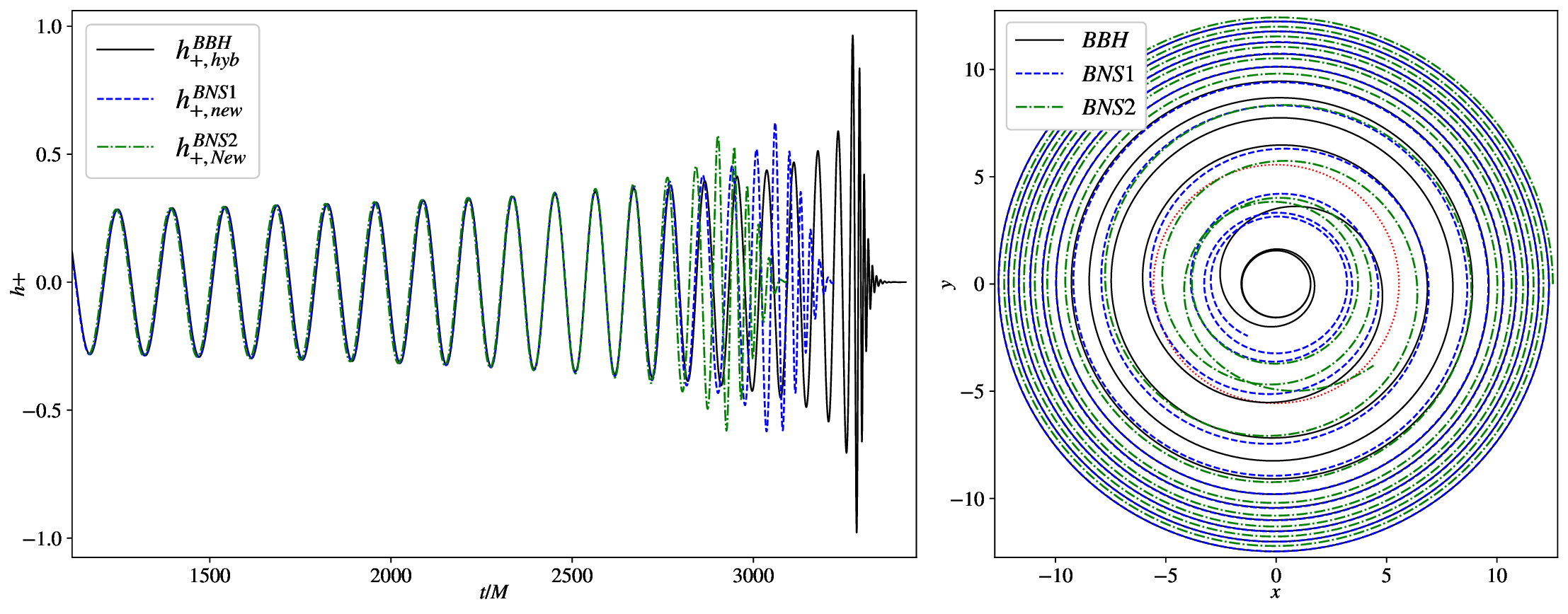}
\caption{Our analytical strain (left plot) and separation (right plot) for the baseline BBH (solid black), the $\Lambda=791$ BNS1 (dotted blue) and the $\Lambda=1520$ BNS2 (dashdot cyan) for the last $\approx 10$ orbits extended to the early stages of the collision. 
The dotted red line marks where the neutron stars touch.}
\label{Figure8}
\end{figure}

We plot in Fig. \ref{Figure8} the $h_{+}$ component of the strain for the baseline model, in comparison with the BNS strain for the two equations of state considered. 
While our model for the amplitude past the merger is simplified, the phase modeling remains accurate.
In the left plot of Fig. \ref{Figure8} we show the evolution of the separation between the neutron stars for the two equations of state considered, continued beyond the merger, depicted by the red dotted circle, in comparison to the separation of the baseline BBH model.
We see how the value of the tidal deformability and thus of the equation of state influence the early stages before and beyond the collision, revealing the effect of the matter interaction on the orbits.
A larger tidal deformability (dashdot cyan) speeds up the merger of the stars, leading to a larger-radius remnant, whereas a smaller value for it (dotted blue) prolongs the inspiral and yields a denser remnant with a smaller radius.

%%================================%%
\section{Conclusion}\label{conclusion}

%We present \texttt{RisingTides}, our new Python code steered by a user-friendly Jupyter notebook, that calculates analytical GW templates %during the merger of compact binary objects (BBH and BNS).
%We tested analytical approximations for the tidal phase and amplitude of BNS and re-calibrated the parameters.
%Our results are relevant for the characterization of compact binary mergers accompanied by EM counterparts.

%The goal is to accurately model the BNS merger physics.
% show that tidal deformability at the merger are tightly correlated with the EOS
%One solution would be to build a hybrid tidal phase correction, where we keep the tidal approximation dependent linearly on the deformability up to ISCO, but. after that use a new approximant, that depends in a more complicated way on the deformability. 
%In the future we will explore universal relations for BNS.
%allows to derive semi-analytical expressions for the

The gravitational waves emitted during a BNS merger offer insights into the behavior of matter under extreme conditions. 
To understand the impact of matter on the evolution of such systems and on their gravitational waves signature, we embarked upon an analytical modeling journey.
Initially, we focused on a BBH system, employing the post-Newtonian formalism for the inspiral and the Backwards-one-Body model for the merger. 
By combining them, we established a baseline waveform, which we validated against numerical relativity simulations, to ensure a robust foundation for our next step. 
To incorporate the effects of tides, we introduced corrections to the phase and amplitude of the point-particle waveform by using the polynomial expressions proposed by the NRTidal model. 
We then verified the model's accuracy and efficiency through careful comparison with numerical relativity data for two equations of state. 

However, we encountered a mismatch around the merger when solely relying on the NRTidal model. 
To address this limitation, we lifted the restriction on the coefficients' independence from tidal deformability and recalibrated them with the numerical relativity predictions.
 By performing new fits to the numerical BNS data, we obtained updated values for the polynomial coefficients.
Armed with these new coefficients, we reconstructed the tidal corrections and achieved improved fits for the phase and amplitude, successfully extending the phase modeling beyond the merger. 
To achieve a comprehensive phase representation, we devised a method to determine its extent, and applied a taper at the end, seamlessly continuing with the baseline model.
Regarding the amplitude modeling, we have successfully carried it up to the merger, where we employed a Hann taper to ensure a smooth and continuous transition into the post-merger.
We ended by reconstructing the complete analytical BNS strain and by investigating the tidal influence on the system's orbits around the BNS collision.

We developed \texttt{RisingTides}, a Python code guided by a user-friendly Jupyter Notebook, for the analytical modeling the tidal effects on the gravitational waves emitted during BNS inspiral and merger.
We make our implementation available to the scientific community, to foster collaboration and facilitate further future investigations.

In our future work, we aim to enhance our model by incorporating a broader range of numerical relativity BNS simulations encompassing diverse equations of state. 
We will seek to uncover a consistent pattern for the dependence of the phase on the tidal deformability around the merger, ultimately leading to a universal set of polynomial coefficients.
Furthermore, we will explore universal relations for BNS systems, which may offer deeper insights into their behavior and characteristics. 
Additionally, we are committed to refining the accuracy of the amplitude modeling beyond the merger, thus increasing the overall predictive power of our approach.
Our research will contribute to a better understanding of BNS mergers and their gravitational wave signatures.

\backmatter

\bmhead{Supplementary information}
We release \texttt{RisingTides}, an open-source Python code for the analytical modeling of tidal effects on gravitational waves from BNS mergers.
Our implementation can be accessed on \texttt{GitHub}, or through \texttt{Zenodo}.

\bmhead{Data Availability}
The datasets generated and analyzed during the current study are openly available at \href{https://github.com/mbabiuc/RisingTides.git}{github.com/mbabiuc/RisingTides}, linked to the open-access \texttt{Zenodo} repository, at \href{https://zenodo.org/badge/latestdoi/671997364}{DOI: 10.5281/zenodo.8206261}.
%Data that supports the findings of this study 
\bmhead{Acknowledgments}

The authors are wishing to acknowledge the Physics Department and the College of Science at Marshall University.
This research was supported in part by the National Science Foundation under Grant No. NSF PHY-1748958.

%%================================%%
%\TD{I AM HERE}
%%================================%%
%%%Power verbs
%%Let us suppose; consider; contrast this with; before we turn to; up to this point we have not; let us take a closer look at; let us now attempt; indeed, we shall see; clearly, what is happening here is that; seeing this, we should not be surprised to find; remember that; it is sometimes helpful to think of it as; going beyond this approximation brings us up against; we now return to; the reader is warned; a cautionary remark; the reader may well ask, will rapidly converge; has long been known; could severely limit; may also be incomplete; would then follow; might incorrectly assume; can no longer be seen; had not yet received.

%\begin{appendices}

%\section{title}\label{appendixA}
%%=============================================================%%
%\end{appendices}

%https://ui.adsabs.harvard.edu
\bibliography{references}% common bib file

\end{document}